\documentclass[a4paper,11pt]{article}
\usepackage{pos}
\usepackage{epsfig,dsfont}
\usepackage{amsmath}
\newcommand{\SU}{\text{SU}}
\newcommand{\SO}{\text{SO}}
\newcommand{\Z}{\mathbb Z}
\newcommand{\U}{\text{U}}
\usepackage{todonotes}
\usepackage{bm}

\title{U(1)-gauged 2-flavor spin system in 3-D}

\author*[a,c]{Christof Gattringer}
\author[b]{Tin Sulejmanpasic}

\affiliation[a]{FWF Austrian Science Fund\\
  Georg-Coch-Platz 2, 1010 Vienna, Austria}

\affiliation[b]{University of Durham\\
Durham, UK}

\affiliation[c]{On leave of absence from:
Institute of Physics, University of Graz, Austria}

\emailAdd{christof.gattringer@fwf.ac.at}
\emailAdd{tin.sulejmanpasic@durham.ac.uk}

\abstract{We study a U(1)-gauged 2-component spin system in 3 dimensions. For the gauge fields we use the Villain formulation with a constraint that removes 
lattice monopoles and in this form couple the gauge fields to 2-component spins. We discuss the simulation strategies for this highly constraint system and present 
first results for the phase structure. Our preliminary Monte Carlo simulations indicate that the system undergoes a second order phase transition driven by 
the spin coupling. However, the correlation length critical exponent is inconsistent with the conformal bootstrap constraint for a critical (rather than multi-critical) fixed 
point, which forces a conclusion that the transition is likely weakly 1st order, passing close to a multi-critical point. To reliably resolve the nature of the transition, 
simulations on larger lattices will be necessary.}

\FullConference{The 41st International Symposium on Lattice Field Theory (LATTICE2024)\\
 28 July - 3 August 2024\\
Liverpool, UK\\}

\begin{document}
\maketitle

\section{Introductory remarks} 

\begin{figure}[t] 
\centering
\includegraphics[width=\textwidth]{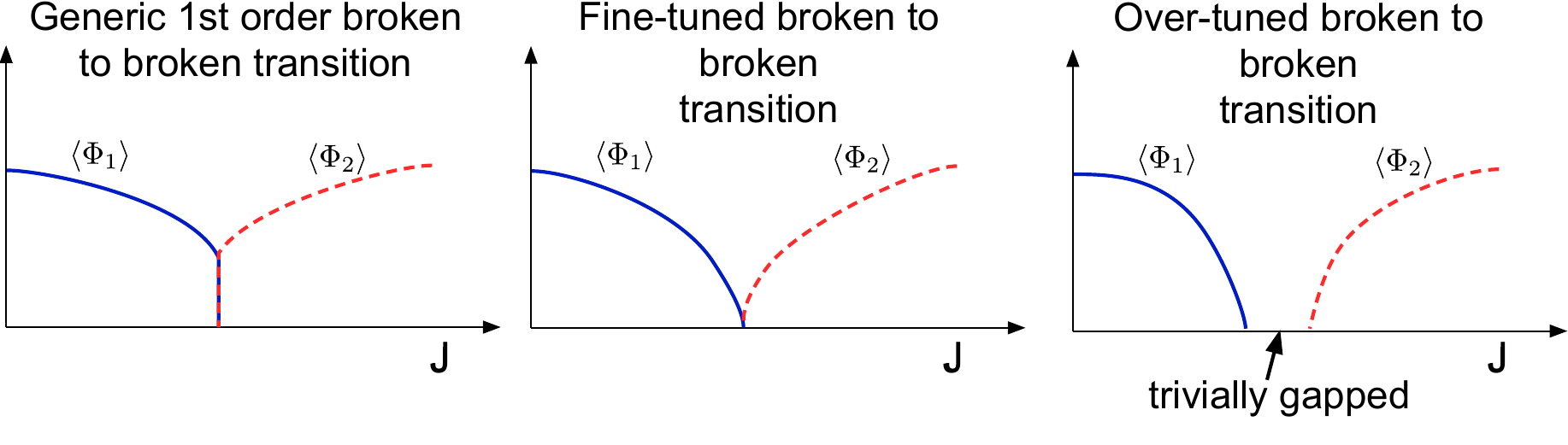} 
\vskip-3mm
\caption{The Landau-Ginzburg expectation for the broken-to-broken transition. $\Phi_{1}$, $\Phi_2$ are the order parameters of the symmetries 
$G_1$, $G_2$, and $J$ is a parameter that is tuned to induce a transition. The generic direct transition from a $G_{1}\times G_2\rightarrow G_2$ to 
$G_{1}\times G_2\rightarrow G_1$ phase is 1st order (left). As one tunes another parameter one can obtain a picture where both order parameters 
go to zero at the same point (middle). As one overtunes this parameter, a trivially gapped phase emerges and the multi-critical 
point splits into two.}
\label{fig:landau}
\vskip-3mm
\end{figure}

\noindent
Phase transitions and symmetries have a close connection. When a system has a particular symmetry, it can typically be driven into a phase where the symmetry is 
spontaneously broken, or restored. According to the Landau-Ginzburg paradigm, the effective description of this transition can be written as an effective quantum 
field theory of the order parameter, and thus, since any two systems with the same symmetries can be captured by such an effective description, all of such 
transitions are universal. 
According to this lore, a transition with a symmetry group of the form $G_1\times G_2$ from a phase $1$ where the symmetry breaking pattern is  
$G_1\times G_2\rightarrow G_1$ to a phase where the symmetry breaking pattern is $G_1\times G_2\rightarrow G_2$ must generically be of 1st order. 
The reason is that, unless there is some fine tuning, there is no reason for both order parameters to go to zero at the same time. 
As a consequence the system is typically multi-critical -- see Fig.~\ref{fig:landau}.

\begin{figure}[htbp] 
\centering
\includegraphics[width=\textwidth]{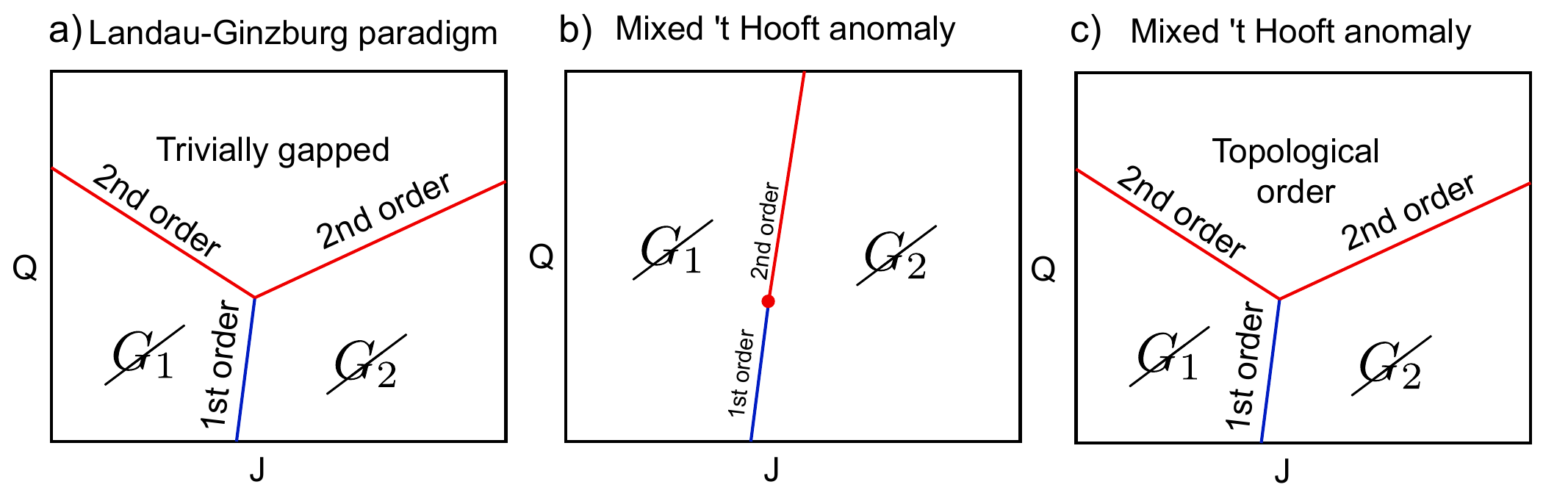} 
\vskip-2mm
\caption{If two parameters $J, Q$ are changed, a generic 1st order order-to-order transition becomes two 2nd order transitions, 
according to Landau-Ginzburg theory a). 
If a mixed 't Hooft anomaly is present, there can be no trivially gapped phase, and so either the transition can become a 2nd order line b) or another 
exotic gapped phase, e.g., with topological order sketched in c).}
\label{fig:landau2}
\vskip-3mm
\end{figure}

However, if the symmetry group has a mixed `t Hooft anomaly\footnote{A mixed 't Hooft 
anomaly can be thought of as 
breaking of the symmetry $G_1$ if gauge fields for the symmetry $G_2$ are introduced.}
between the symmetries $G_1$ and $G_2$ , the Landau-Ginzburg 
paradigm breaks down. A mixed 't Hooft anomaly requires that either $G_1$ or $G_2$ is spontaneously broken, and so the trivially gapped phase is impossible, i.e., 
the scenario in the far left of Fig.~\ref{fig:landau} cannot happen. Thus it is plausible that a generic 2nd order order-to-order transition can happen in systems 
with a mixed 't Hooft anomaly (see Fig.~\ref{fig:landau2}b). However, also 
another gapped phase could emerge, potentially developing some kind of topological order.

The most prominent system which was suggested to have a generic 2nd order transition is that of a 2+1d spin-1/2 anti-ferromagnet, going from a N\'eel to the VBS 
phase, for which a generic 2nd order transition was proposed to exist \cite{Senthil:2003eed}. The N\'eel phase is a phase where the $\SO(3)$ spin symmetry\footnote{The symmetry 
of spin $1/2$ is $\SO(3)$, not $\SU(2)$ because the spin operator $\bm S=(S_1,S_2,S_3)$ is an $\SO(3)$ vector, and there exists no operator which transforms in 
an $\SU(2)$ representation which is not in $\SO(3)$. } is spontaneously broken to $U(1)$, while the VBS phase is a phase where two neighboring spins combine 
into a singlet dimer and spontaneously break the lattice symmetries. The symmetry breaking pattern for the VBS phase depends on the details of the lattice, but for 
a square lattice the breaking leaves $4$ vacua related by the group $\Z_4$ generated by the 90-degree lattice rotations. 

The basic premise is that the spin model in question is effectively described by scalar QED in 3d, with a gauge-charge unity $\SU(2)$ scalar doublet 
$\Phi = (\phi_1, \phi_2)$. The model has an $\SO(3)$ flavor symmetry rotating the scalars\footnote{Note that because the scalar doublet carries gauge 
charge one, the action of the center of $\SU(2)$ is given by $\Phi\rightarrow -\Phi$, which is a gauge symmetry, not a global symmetry. Hence the global symmetry 
is $\SO(3)$, not $\SU(2)$. }, i.e., $\Phi\rightarrow U\Phi, \, U\in\SU(2)$, consistent with the spin-symmetry of the spin-1/2 anti-ferromagnet. Further, since the 
system has an abelian gauge field $A_\mu$, it also has a conserved current $j^\mu=\frac{1}{2\pi}\epsilon^{\mu\nu\rho}\partial_\nu A_\rho$, where $
\epsilon^{\mu\nu\rho}$ is the totally anti-symmetric tensor. The validity 
of $\partial_\mu j^\mu=0$ is a statement that there exists a $\U(1)$-magnetic symmetry, the 
so-called magnetic symmetry, which is not a symmetry of the spin-1/2 antiferromagnet. This symmetry is broken by the presence of magnetic monopole operators in 
the theory. However, a relevant symmetry is a $\mathbb Z_4$-symmetry of lattice rotations by 90 degrees, which in the infrared theory 
transmutes\footnote{It may come as a surprise that the 
spatial lattice symmetry has anything to do with the internal symmetry of the effective model. The statement is that an operator which preserves the symmetry of the 
lattice, but breaks the $\U(1)$ symmetry completely has to have a large momentum, of the order of the inverse lattice system size, and as such decouples in the 
thermodynamic limit. Such symmetries are dubbed emanent symmetries \cite{Cheng:2022sgb}.} into the $\mathbb Z_4$-subgroup of the $
\U(1)$-magnetic symmetry. Hence an effective theory of the spin-1/2 system is the 2-scalar QED with charge $4$ monopoles, 
preserving the $\Z_4$-subgroup of $\U(1)$-magnetic.

The presence of a generic 2nd order transition requires that the charge-$4$ monopole operators are irrelevant at the corresponding conformal fixed 
point\footnote{The fixed point is sometimes called deconfined critical point (DCP).}, which is indeed the case in the $SU(N)$ generalizations of the model when $N$ 
is large \cite{Murthy:1989ps,Senthil:2003eed}. However, the generic 2nd order nature has a controversial history (see \cite{Takahashi:2024xxd} for a recent paper). Recent results on the spin-1/2 
models seem to indicate that the scenario Fig.~\ref{fig:landau2}c) is realized. However, in spin systems at least the $\U(1)$-magentic symmetry needs to 
emerge\footnote{It is believed that perhaps even $\SO(5)$ symmetry emerges at the (multi-)critical point. See \cite{Takahashi:2024xxd} and references therein.}. In this work instead, we 
consider the lattice formulation of the 2-scalar QED model with an exact magnetic symmetry\footnote{Note that in \cite{Liu:2018sww} Fermionic in the same universality class with the exact $\U(1)$ symmetry was also considered.} which is possible using the modified Villain formalism\footnote{We would like to point out some interesting development in the direction of the Villainized modal of non-abelian gauge theory \cite{Chen:2024ddr,Zhang:2024cjb}. } \cite{Sulejmanpasic:2019ytl,Anosova:2019quw,Gorantla:2021svj,Anosova:2022cjm,Anosova:2022yqx,Fazza:2022fss,Jacobson:2023cmr,Nguyen:2024ikq,Jacobson:2024hov,Xu:2024hyo,Peng:2024xbl}. Preliminary results presented here seem to indicate a 2nd order phase transition, but further simulations on 
larger lattices are needed to reliably resolve the nature of the transition. 

\section{Definition of the model}

\noindent
The system we study is defined on a 3d lattice with volume $V = L^3$. All fields
obey periodic boundary conditions. The partition function $Z$ depends on two couplings, the inverse gauge coupling $\beta$ and the matter coupling $J$,
\begin{equation}
Z \, = \; \int \!\!D[A] \,   B_\beta[A] \, \int \!\!  D\big[ \Phi\big] \, e^{\, J \, S_{spin} [A,\Phi]} \; .
\label{eq:Zdef}
\end{equation}
The gauge field degrees of freedom are $A_{x,\mu} \in [-\pi,\pi]$ assigned to the links of the lattice, with the corresponding 
path integral measure given by $\int \!\!D[A] = \prod_{x,\mu} \int_{-\pi}^\pi \! dA_{x,\mu}$.
The Boltzmann factor $B_\beta[A]$ for the gauge degrees of freedom is given by 
\begin{eqnarray}
B_\beta[A] & = & \sum_{\{k\}} \, e^{ \, - \, \beta \, S_{gauge}[A,k]}  \; 
\prod_x \delta \big( (dk)_{x,123} \big) \;  , \; \; 
\label{Boltzmann_A}
\\
S_{gauge}[A,k] & = & \frac{1}{2} \sum_{x,\mu<\nu} \big( (dA)_{x,\mu \nu} \, + \, 2\pi k_{x,\mu \nu} \big)^2 \; ,
\label{Sgauge}
\end{eqnarray}
where $\sum_{\{k\}}$ is the sum over all configurations of the plaquette-based Villain variables $k_{x,\mu \nu} \in \mathds{Z}$, defined as
$\sum_{\{k\}} = \prod_{x,\mu < \nu} \sum_{k_{x,\mu \nu} \in \mathds{Z}}$, and
$(dA)_{x,\mu \nu} = A_{x+\hat\mu,\nu} - A_{x,\nu} - A_{x+\hat\nu,\mu} + A_{x,\mu}$ denotes the exterior derivative of the gauge fields. 
The Villain variables are subject to constraints that in (\ref{Boltzmann_A}) are implemented with a product of Kronecker deltas 
(here denoted as $\delta(n) = \delta_{n,0}$) that enforce
\begin{equation}
(dk)_{x,123} \; = \; 0 \; \forall x \quad \mbox{with} \quad (dk)_{x,123} \, = \, 
k_{x+\hat3,12} \, - \, k_{x,12} \, - \, k_{x+\hat2,13} \, + \, k_{x,13}  \, + \, k_{x+\hat1,23} \, - \, k_{x,23} \; , 
\label{constraint}
\end{equation} 
which implies the absence of monopoles. The spin degrees of freedom are given 
by\
\begin{equation}
\Phi_x \;  = \;  \left(
\begin{array}{c}
\cos \theta_x \; e^{i \alpha_x} \\
\sin \theta_x \; e^{i \beta_x} 
\end{array}
\right) \quad \mbox{with} \quad \theta_x \in [0,\pi/2] \, , \; \; \alpha_x, \beta_x \in [-\pi, \pi] \; .
\end{equation}
The path integral measure is again a product measure,
$D\big[ \Phi\big]  = \prod_x \int_{0}^{\pi/2} \!\! d \theta_x \int_{-\pi}^{\pi} \! d \alpha_x \int_{-\pi}^{\pi} \! d \beta_x$.
Finally, the action for the spin degrees of freedom is
\begin{eqnarray}
&& S_{spin} [A,\Phi]  =  \frac{1}{2} \sum_{x,\mu} \big[ \Phi_x^\dagger \, e^{\, i \, A_{x,\mu} }\, \Phi_{x + \hat \mu} \; + \; c.c.\big] \\ 
&&  \quad = 
\sum_{x,\mu} \! \big[ \cos \theta_x \, \cos \theta_{x + \hat \mu} \, \cos(A_{x,\mu} + \alpha_{x + \hat \mu} - \alpha_x) \, + \,
\sin \theta_x \, \sin \theta_{x + \hat \mu} \, \cos(A_{x,\mu} + \beta_{x + \hat \mu} - \beta_x) \big] .
\nonumber
\end{eqnarray}

\section{Observables}

\noindent
For our analysis we use two order parameters. The spin order parameter (magnetization) is defined as, ($\vec \sigma$ is the vector of Pauli matrices)
\begin{equation}
M \; = \;  \Big| \sum_{x} \vec{S}_x \, \Big|  \; , \quad \mbox{with} \quad  \vec{S}_x \; = \; \Phi_x^\dagger \, \vec{\sigma}/2 \,  \Phi_x \; ,
\label{orderparamM}
\end{equation}
with a non-zero expectation value of $M$ signalling the breaking of SO(3). Our second order parameter is the total net winding number of closed monopole loops
on the dual lattice,
\begin{equation}
\omega \; = \; \frac{1}{3L^2} \left( \; \bigg| \sum_{x:\, x_1 = 0} k_{x,23} \bigg| \, + \, 
\bigg| \sum_{x:\, x_2 = 0} k_{x,13} \bigg| \, + \, \bigg| \sum_{x:\, x_3 = 0} k_{x,12} \bigg| \; \right)  .
\label{Omega}
\end{equation} 
When viewed on the dual lattice, where we denote sites as $\tilde x$ and the dual Villain variables as $\tilde k_{\tilde x,\nu}$,
it is obvious that the sum $\sum_{x:\, x_1 = 0} k_{x,23}$ turns into
$\sum_{\tilde x:\, \tilde x_1 = 0} \tilde k_{\tilde x,1}$, which corresponds 
to the total net flux of $\tilde k_{\tilde x,1}$ through the dual 2-3 plane at $\tilde x_1 = 0$ (thus the normalization with $L^2$ to obtain an intensive quantity).
On the dual lattice the constraint (\ref{constraint}) turns into the zero divergence condition $\sum_\mu [\tilde k_{\tilde x,\mu} - \tilde k_{\tilde x - \hat \mu,\mu}] = 0$,
which implies flux conservation of the $\tilde k_{\tilde x,\mu}$. As a consequence $\sum_{\tilde x:\, \tilde x_1 = 0} \tilde k_{\tilde x,1}$ 
equals the total net winding number around the periodic 1-direction, and similarly 
the other two sums correspond to winding around the 2- and 3-directions. The observable thus gives the absolute value of the net 
flux density averaged over all three directions of the dual lattice. Note that the constraint (\ref{constraint}) forbids isolated monopoles or anti-monopoles, 
but a non-zero winding number of closed monopole loops on the dual lattice is admissible. This is what $\omega$ is measuring. 

For locating the critical point and a first analysis of its universality class we furthermore
study the susceptibility $\chi_M$ of $M$ and the corresponding Binder cumulant $U_M$,
\begin{equation}
\chi_{\, M} \; = \; \frac{1}{L^3} \; \left\langle \Big( M \, - \, \langle M \rangle \Big)^2 \right\rangle \; , \; \;
U_M \; = \; 1 \; - \; \frac{3}{5} \, R_{42} \quad \mbox{with} \quad R_{42} \; = \; \frac{\langle M^4 \rangle}{\langle M^2 \rangle^2} \; , \; 
R_{21} \; = \; \frac{\langle M^2 \rangle}{\langle M \rangle^2} \; ,
\label{binder}
\end{equation}
where for later use we also defined the Binder ratios $R_{42}$ and $R_{21}$. Here we have a 3-component vector as 
order parameter such that the normalization factor in the Binder cumulant is $3/5$.  

\section{Numerical simulation}

\noindent
So far only local updates are implemented for all degrees of freedom, but a more advanced update strategy could use worms for the Villain variables 
$k_{x,\mu \nu}$. For the angles $\alpha_x \in [-\pi, \pi]$ we use proposal values $\alpha_x^\prime = \alpha_x + \Delta$ and in case of 
$\alpha_x^\prime \notin [-\pi,\pi]$ add $\pm 2 \pi$ in oder to project back into $[-\pi,\pi]$ (periodic completion). 
The update of the angles $\beta_x$ is done in the same way, and in both cases the proposal is accepted or rejected with a Metropolis step.
For the angles $\theta_x \in [0,\pi/2]$ we use proposal values $\theta_x^\prime = \theta_x + \Delta$ and in case of 
$\theta_x^\prime \notin [0,\pi/2]$ we reject the proposal and go to the next spin (hard boundary rejection). If the proposal is admissible,
a Metropolis step is used to accept or reject the proposal. 
For the update of the gauge fields $A_{x,\mu} \in [-\pi, \pi]$ we again use $A_{x,\mu}^\prime = A_{x,\mu} + \Delta$ and project back into $[-\pi,\pi]$
by adding $\pm 2 \pi$ in case $A_{x,\mu}^\prime \notin [-\pi, \pi]$. 

In the update of the Villain variables $k_{x,\mu \nu}$ the constraints (\ref{constraint}) 
must be taken into account. For this purpose we use two types of updates: 1.) updates where we 
simultaneously change the Villain variables on four plaquettes that share a common link (i.e., a plaquette on the dual lattice), and 2.) updates where we change 
the Villain variables on stacks of plaquettes. For the dual plaquette update we choose $\Delta \in \pm 1$ randomly and for the Villain variables 
on plaquettes that share the link $(x,1)$ offer the change
$k_{x,12} \rightarrow k_{x,12} + \Delta, \,
k_{x-\hat{2},12} \rightarrow k_{x-\hat{2},12}- \Delta, \,
k_{x,13} \, \rightarrow k_{x,13} + \Delta, \, 
k_{x-\hat{3},13} \rightarrow k_{x-\hat{3},13} - \Delta,$
which we accept or reject with a Metropolis step. In a similar way we change the Villain variables on plaquettes that share links in the 2- and 3-directions. 
It is easy to see that with the correct distribution of signs in the changes $\pm \Delta$ the constraints (\ref{constraint}) remain intact. 
In order to obtain an ergodic update, the dual plaquette update has to be combined with the update of Villain variables on stacks of plaquettes, which correspond to
winding loops on the dual lattice. For example we offer the collective change of Villain variables
$k_{x,12} \rightarrow k_{x,12} + \Delta \; \forall x = (x_1,x_2,x_3) \; \mbox{with} \; x_1, x_2 \; \mbox{fixed, and} 
\; x_3 \, = \, 0,1, \, ... \,  N_3$,
where again $\Delta \in \pm 1$. Again the proposal is accepted or rejected with a Metropolis step.

In this preliminary study we work at an inverse gauge coupling of $\beta = 0.2$ with lattice extents of $L = 4,6,8$ and $L = 10$. We typically use
ensembles with $10^6$ configurations ($5 \times 10^6$ for the Binder cumulants) separated by 50 blocks of updates, where each block combines 5 sweeps through all d.o.f.s 
$\alpha_x, \beta_x, \theta_x, A_{x,\mu}$ and $k_{x,\mu\nu}$ with both, 5 sweeps of local cube updates and 5 dual winding sweeps for the $k_{x,\mu\nu}$. 
We equilibrate the system with $2 \times 10^6$ block updates. All errors we show are 
statistical errors determined with a jackknife analysis combined with blocking of the data.

\section{Results}

\begin{figure}[t]
\centering
\includegraphics[height=60mm]{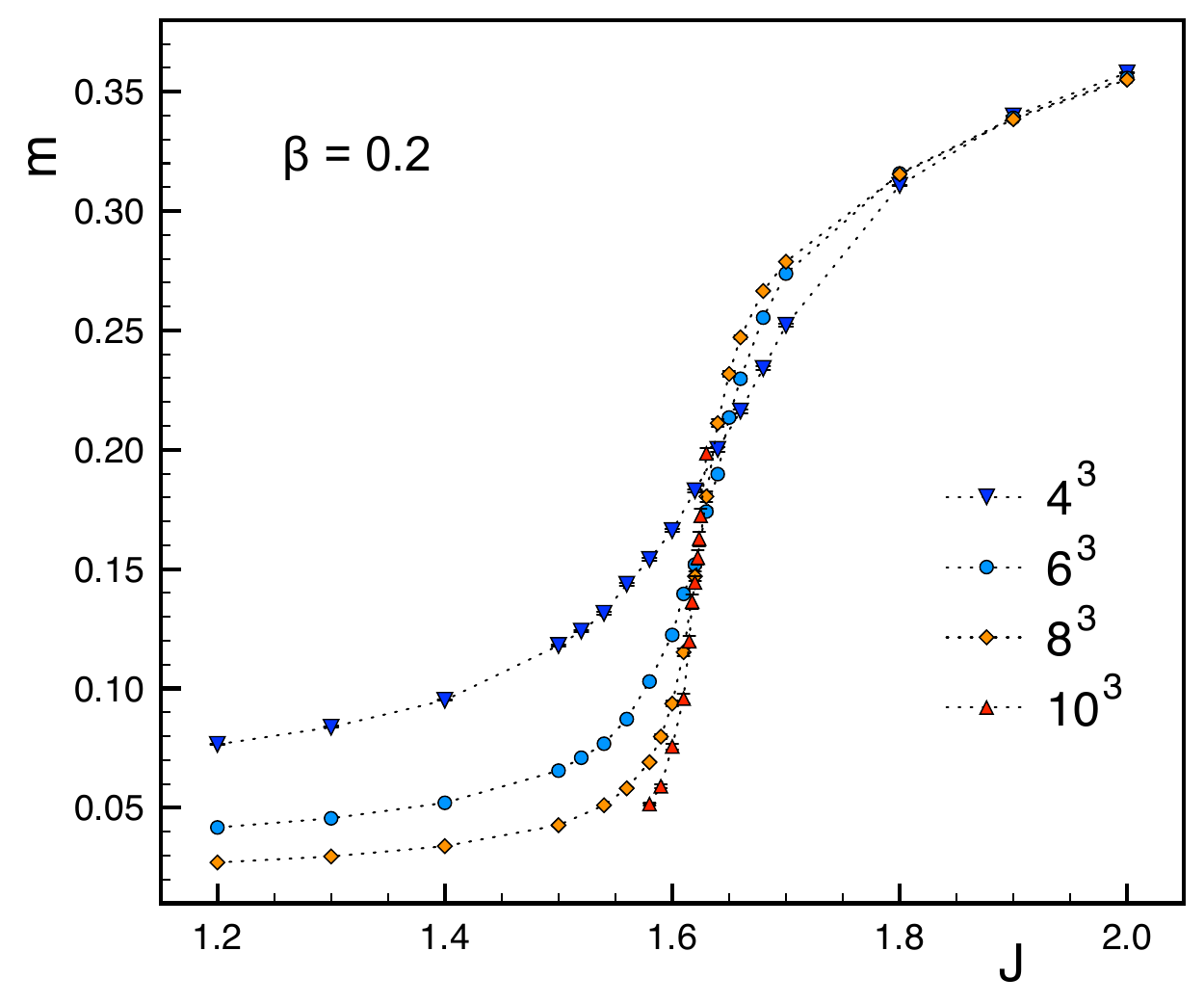}
\includegraphics[height=60mm]{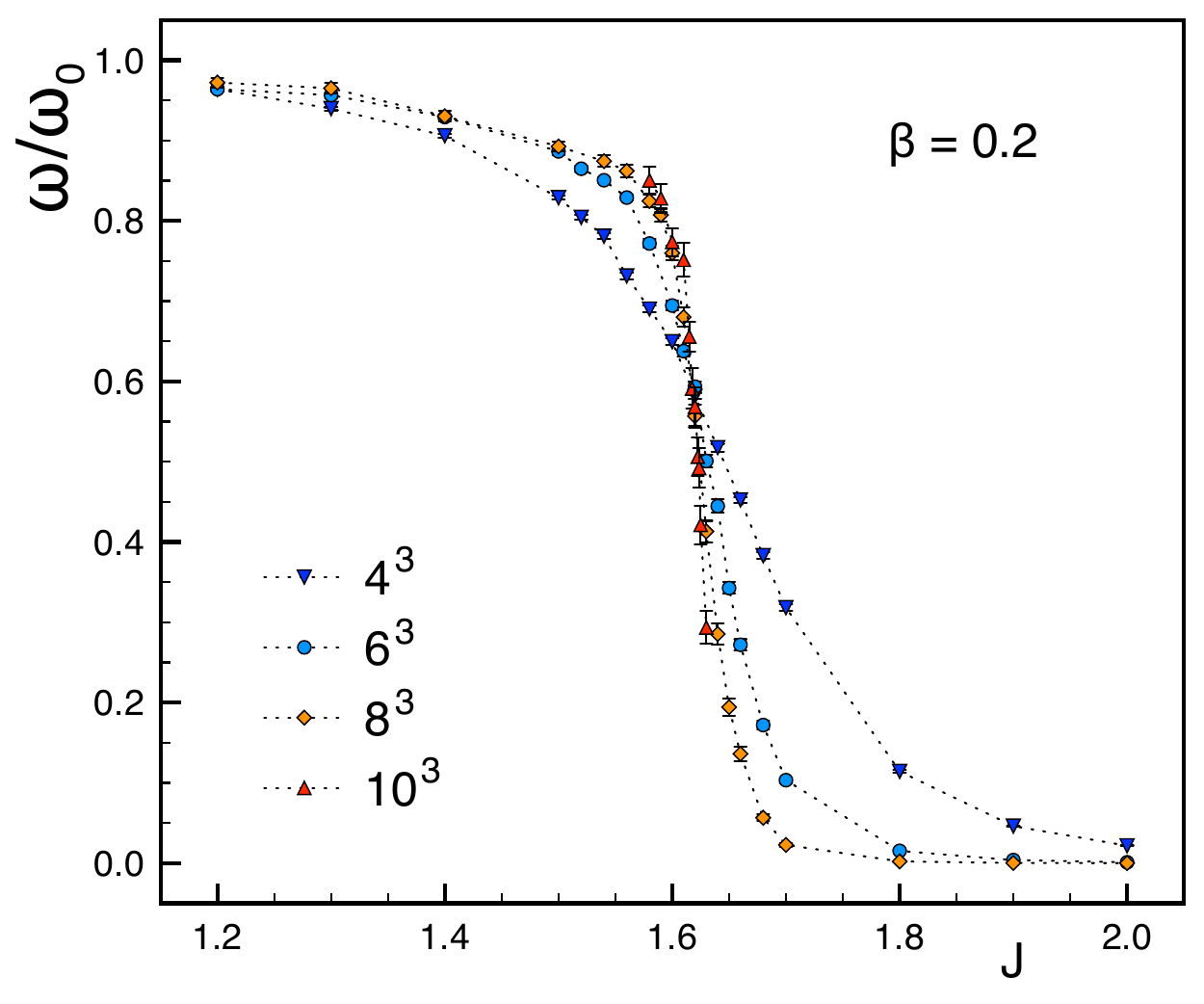}
\vspace{-0.2cm}
\caption{The matter order parameter $m = \langle M \rangle/V$ (lhs.\ plot) and the normalized winding number density $\omega/\omega_0$ (rhs.).
Results are shown as a function of $J$ at $\beta = 0.2$ and we compare data for four different volumes.}
\label{fig:orderparams}
\end{figure}

\noindent
We begin the presentation of our results with the discussion of the order parameters. In the lhs.\ plot of Fig.~\ref{fig:orderparams} we show the 
magnetization density $m = \langle M \rangle/V$ as a function of $J$ for different volumes. It is obvious that at a critical value of $J_c \sim 1.6$ 
the system develops a magnetization and changes from a symmetric phase into a phase where SO(3) is broken.

In the rhs.\ plot of Fig.~\ref{fig:orderparams} we show the winding number $\omega/\omega_0$ where the normalization is done with 
$\omega_0 \equiv \omega \! \mid_{J = 0}$, i.e., the winding number density evaluated at vanishing coupling $J = 0$. Here we observe that at the same critical 
value $J_c \sim 1.6$ the winding number of monopole lines around the periodic boundary conditions drops to zero, while for $J < J_c$ it is finite and 
monopoles could propagate. 
We also analyzed the correlator of a monopole-antimonopole pair which we can generate at the endpoints of a string of dual links along which the corresponding 
Villain variables are changed by $+1$, thus explicitly violating the constraint (\ref{constraint}) with a source term. We find that at $J_c$ the correlator 
abruptly changes its behavior towards fast decay. A more detailed analysis of this behavior is in preparation, but the qualitative study we did so far indicates that the 
drop of the winding number observed in the rhs.\ plot of Fig.~\ref{fig:orderparams}  is due to a change of the monopole-antimonopole correlator at $J_c$. 

\begin{figure}[t]
\centering
\includegraphics[height=60mm]{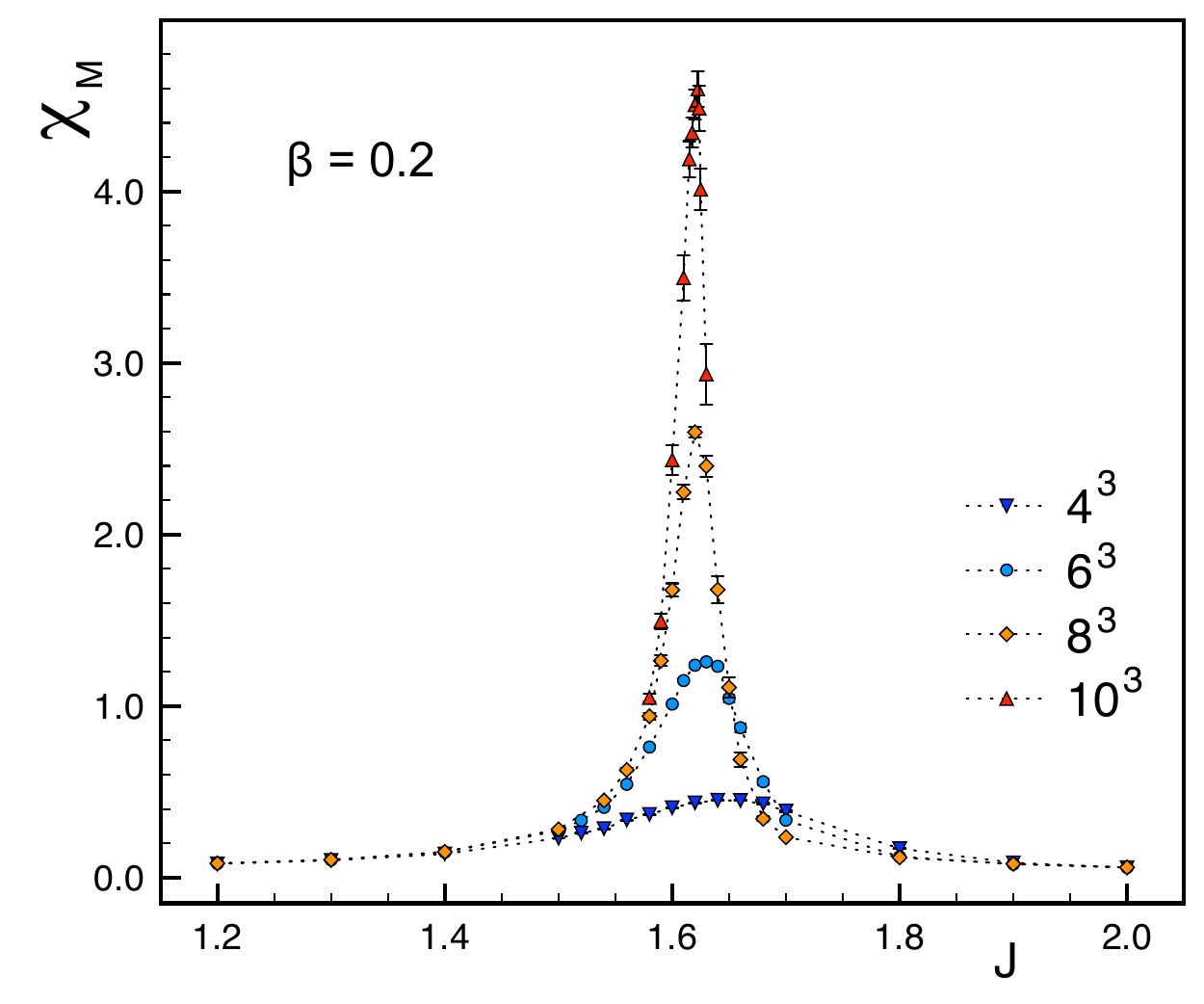}
\includegraphics[height=60mm]{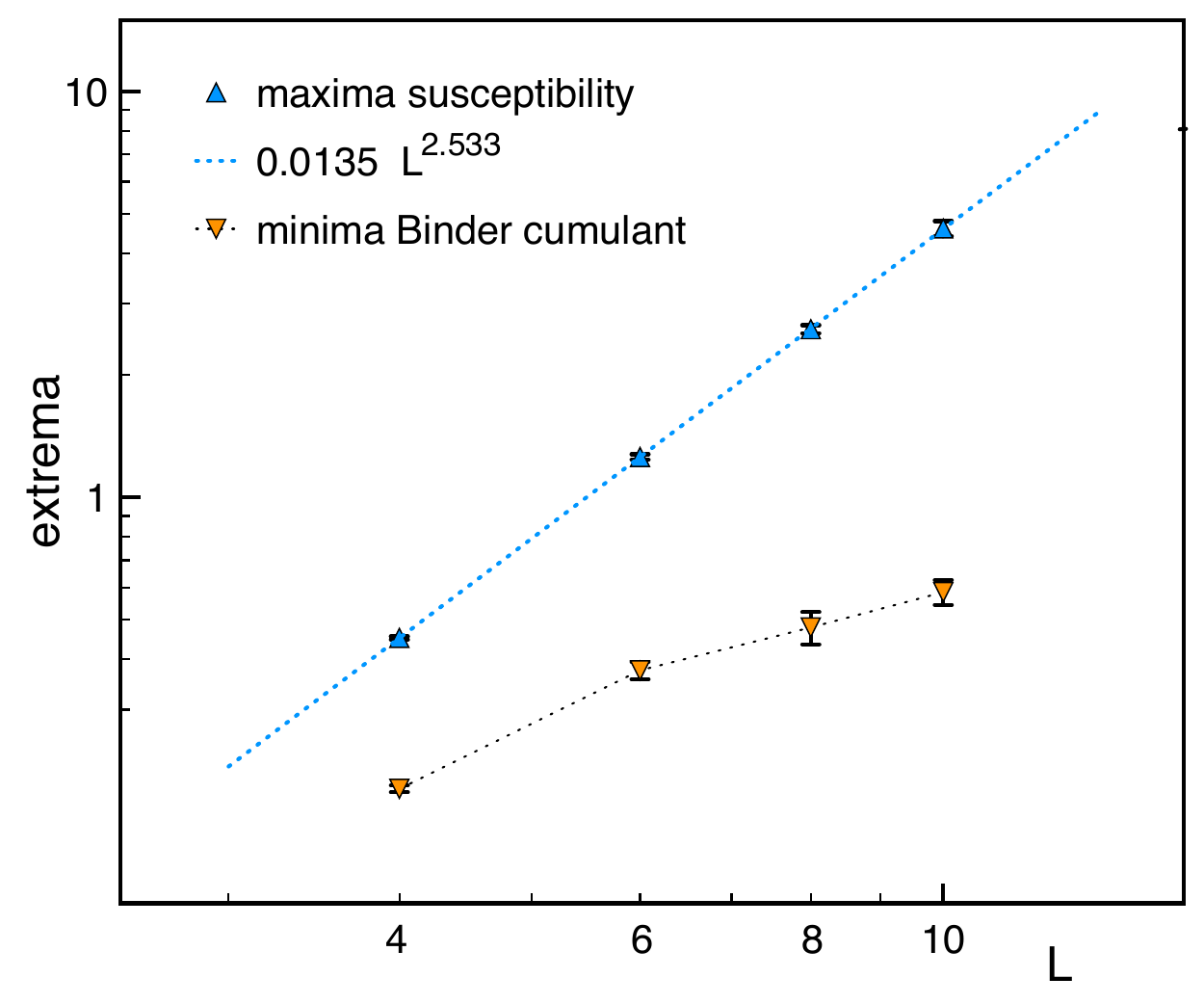}
\vspace{-0.2cm}
\caption{Lhs: Susceptibility $\chi_M$ of the matter order parameter as a function of $J$ at $\beta = 0.2$. 
Rhs: Log-log plot of the maxima of the susceptibility (blue triangles) and the minima of the Binder cumulant $U_M$ (orange triangles) versus the lattice extent 
$L$. The maxima are well described by a power law with exponent 2.533, while for the minima of $U_M$ no such behavior is seen (at least 
not for the small lattices analyzed so far).}
\label{fig:chi_extrema}
\end{figure}

To further characterize the transition at $J_c \sim 1.6$ we study the susceptibility $\chi_M$, which we show as a function of $J$ in the lhs.\ plot of 
Fig.~\ref{fig:chi_extrema}. The susceptibilities develop pronounced peaks that grow with the volume, as is characteristic for a phase transition. 
In the rhs.\ plot we show with blue triangles
the maxima of $\chi_M$ as a function of $L$ in a log-log plot. The growth is well described by a power law with an exponent $\gamma/\nu \sim 2.533$, which 
is smaller than the dimension $d = 3$, thus indicating a continuous transition ($\gamma$ and $\nu$ are the critical exponents for susceptibility 
and correlation length). 

\begin{figure}[t]
\centering
\includegraphics[height=60mm]{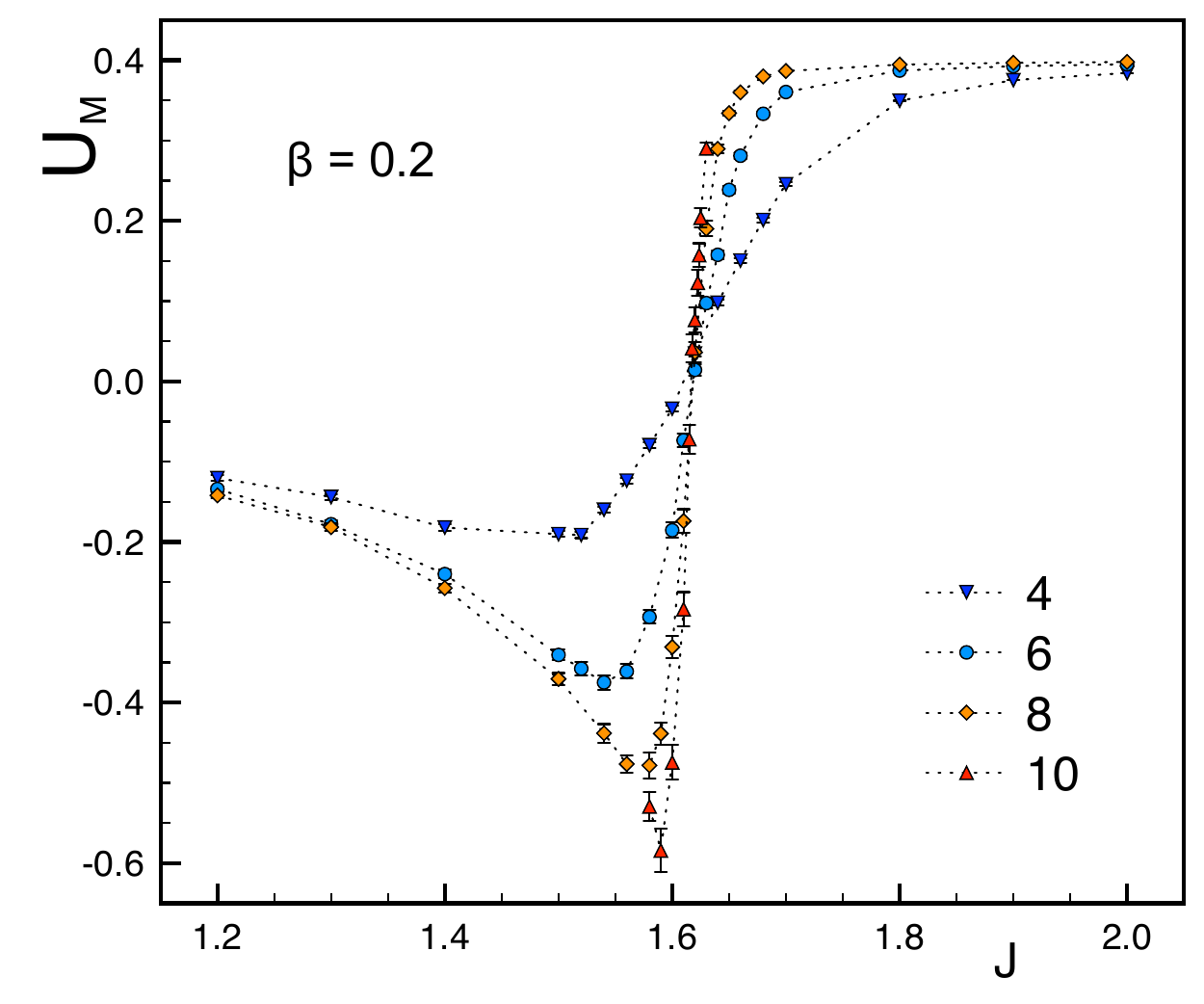}
\includegraphics[height=60mm]{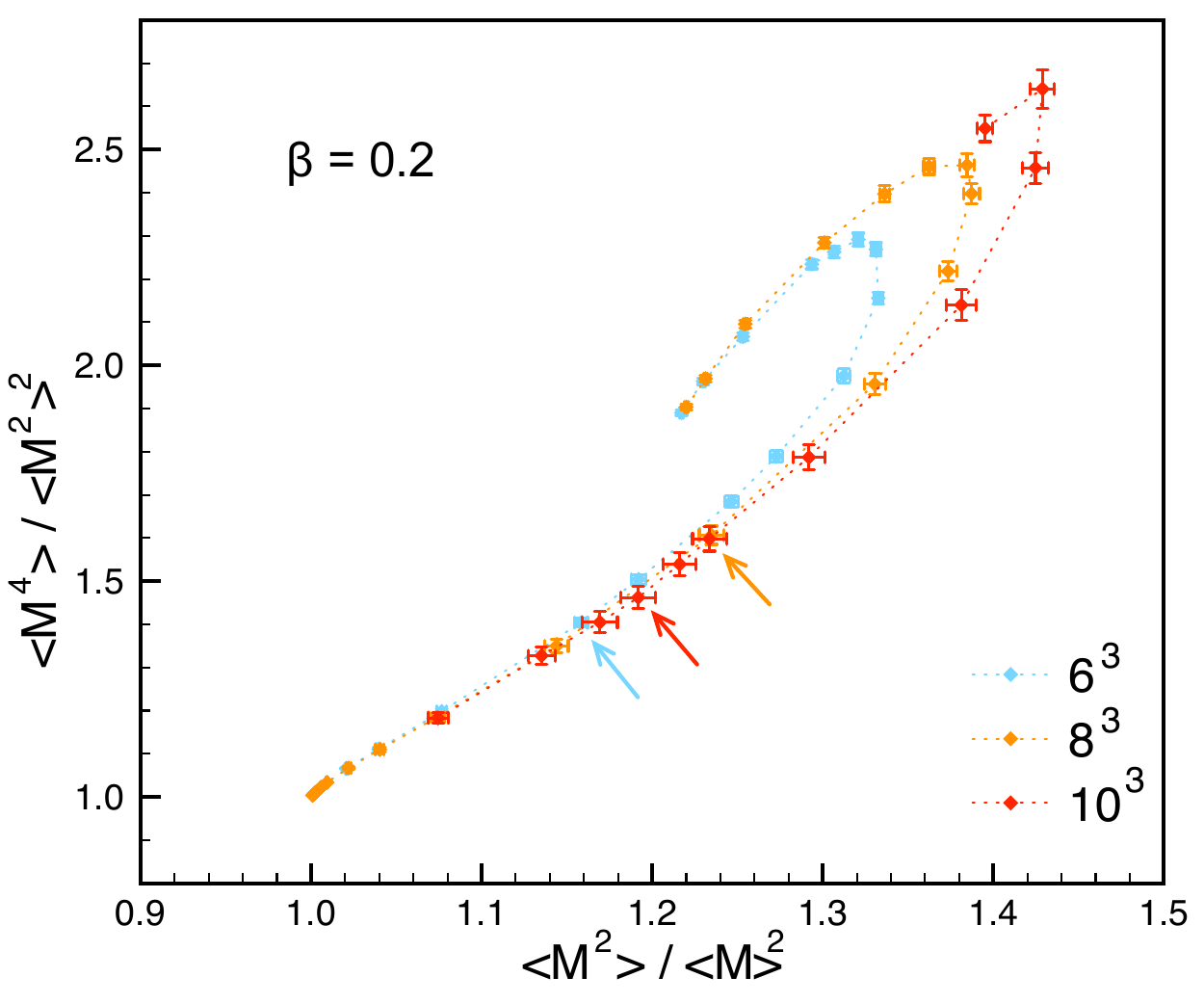}
\vspace{-0.2cm}
\caption{ 
Lhs: The Binder cumulant $U_M$ for different volumes as a function of the coupling $J$ at $\beta = 0.2$.
Rhs: Plot of the Binder ratio $\langle M^4 \rangle / \langle M^2 \rangle^2$ versus the ratio  
$\langle M^2 \rangle / \langle M \rangle^2$ for three different volumes. The arrows show the data points that correspond to the maxima of $\chi_M$, i.e., the
position of the quasi-critical point for the respective volume. This serves to indicate the region where we expect scaling behavior.}
\label{fig:binder1}
\vskip1mm
\includegraphics[height=60mm]{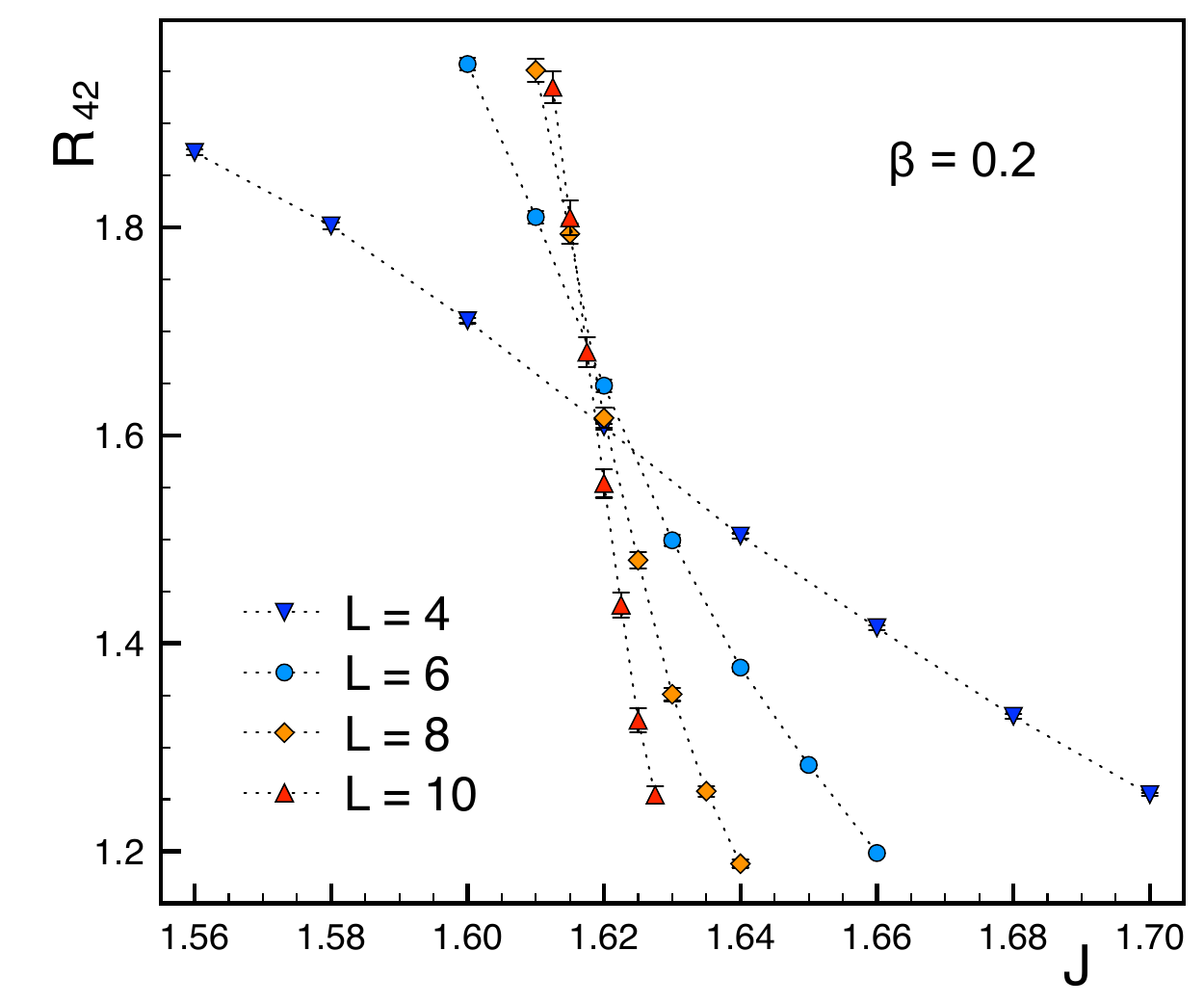}
\includegraphics[height=60mm]{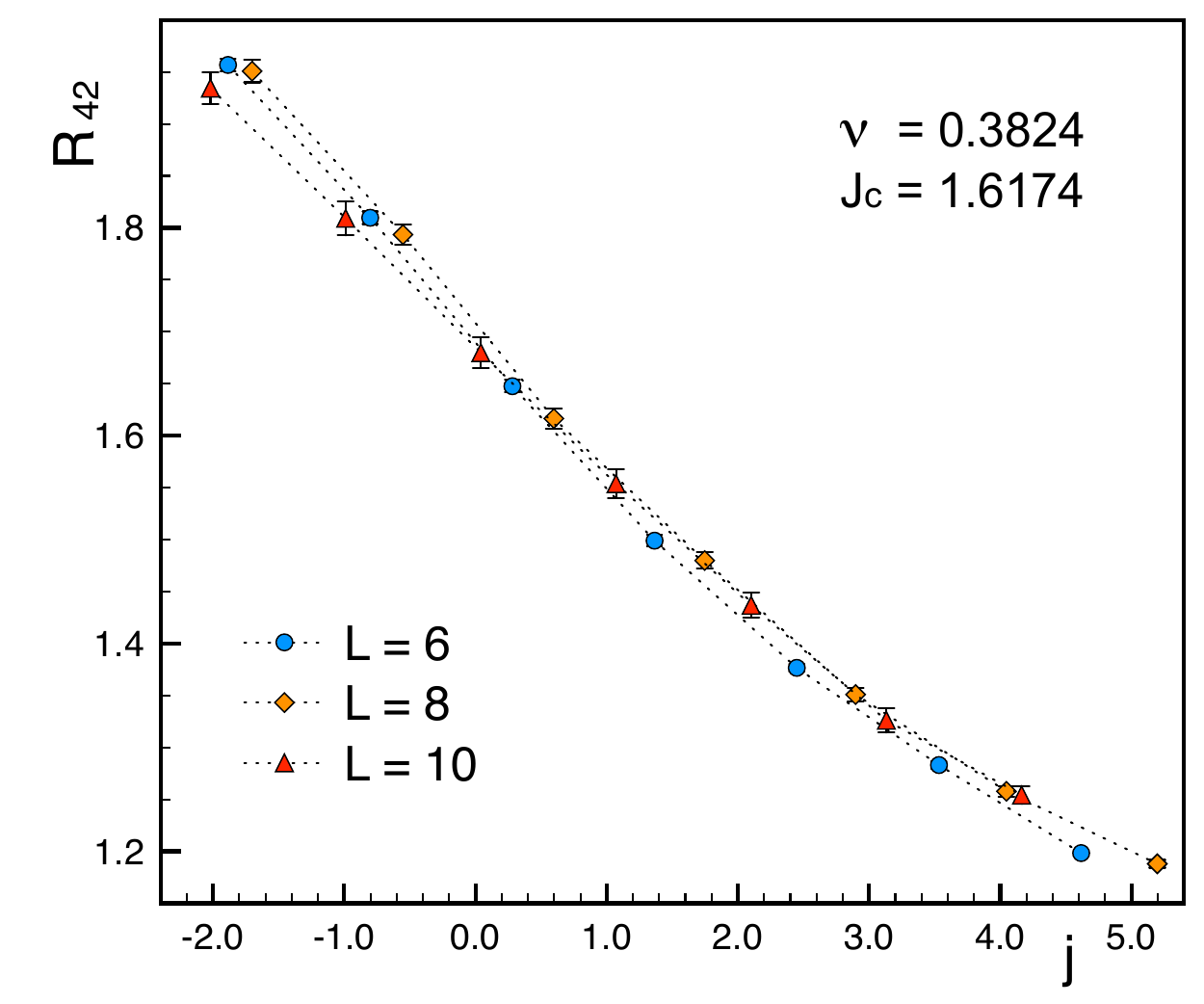}
\vspace{-0.2cm}
\caption{The Binder ratio $R_{42} = \langle M^4 \rangle/\langle M^2\rangle^2$ versus the coupling $J$ (lhs.\ plot) and versus the reduced coupling 
$j = (J-J_c)L^{1/v}$. Results are shown for three different volumes and the optimal collapse of the data in the rhs.\ plot 
is obtained for $\nu = 0.389$ and $J_c = 1.6185$.}
\label{fig:binder_ratios}
\end{figure}
%

We finally come to the analysis of the Binder cumulant. In the lhs.\ plot of Fig.~\ref{fig:binder1} we show the Binder cumulant $U_M$ as defined in (\ref{binder})
as a function of $J$. The Binder cumulant shows the expected crossing of the curves for different volumes at the critical value $J_c \sim 1.6$. However,
the curves also develop minima below $J = 1.6$ which need to be understood, since minima that grow with the volume are an indication for 
first order behavior. In the log-log plot in the rhs.\ of Fig.~\ref{fig:chi_extrema} we also show the size of the minima we observe in the Binder cumulant. Obviously 
we do not see volume scaling, indicating that the minima in the Binder cumulant do not signify a first order transition. 

We remark, that we also conducted a small study in the 3d O(4) model, which our system reduces to when the gauge fields are omitted. 
In the ungauged O(4) model one can directly use the spin expectation value $\langle \Phi \rangle$ as order parameter 
(which in our model is not a gauge invariant observable), 
and based on that analysis it is well established that the O(4) model undergoes a single continuous transition. 
However, it is possible to study this transition also with 
our order parameter $M$ as defined in (\ref{orderparamM}) and also in the O(4) model the corresponding Binder cumulants $U_M$ develop the same minima 
as observed here. The deeper reason for this behavior is that $\langle M^4 \rangle$ is a monomial of order 8 in the spin degrees of freedom, leading to additional 
minima in the observable, which, however, are not linked to some first order behavior. 

Yet another evidence that the transition we observe at $J_c \sim 1.6$ is continuous is presented in the rhs.\ plot of Fig.~\ref{fig:binder1}, where we show
the Binder ratios $\langle M^4 \rangle / \langle M^2 \rangle^2$ versus $\langle M^2 \rangle / \langle M \rangle^2$. According to the finite size scaling ansatz 
for continuous transitions, both these ratios should behave like universal functions $g_4(j)$, $g_2(j )$, 
where $j = (J - J_c)L^{1/\nu}$ and $\nu$ is again the critical exponent of the correlation length. Thus for a continuous transition the curves for the different 
volumes should fall on top of each other in the scaling region (arrows in the plot), and this is what we indeed observe.

To complete our analysis by exploiting the finite size scaling relation for the Binder cumulant $R = \langle M^4 \rangle / \langle M^2 \rangle^2 \sim g_4(j)$ and 
show $R$ as a function of $j = (J - J_c)L^{1/\nu}$ in the rhs.\ plot of Fig.~\ref{fig:binder_ratios} (the lhs.\ plot shows the raw data as a function of $J$). 
When choosing $J_c = 1.6174$ and $\nu = 0.3824$ we find the optimal collapse of the data. Repeating the collapse with the Binder cumulant 
$R_{21}$, we find $J_c = 1.6191$ and $\nu = 0.3721$. 

However, we remark that $\nu=0.3721$ is inconsistent with the conformal bootstrap constraint for having a critical, rather than multi-critical, point (see \cite{Nakayama:2016jhq} Appendix C). While the analysis is still preliminary, this indicates that the transition is likely weakly first order, passing near a multi-critical fixed point, supporting scenario  Fig.~\ref{fig:landau}c. Only simulations on larger volumes will resolve this issue.

\acknowledgments

We thank Fakher Assad for discussions at SIGN-25. TS is supported by the University Research Fellowship of the Royal Society of London and, in part, by the STFC grant number ST/T000708/1. 

\bibliographystyle{JHEP} 
\bibliography{bibliography} 

\providecommand{\href}[2]{#2}\begingroup\raggedright\begin{thebibliography}{10}

\bibitem{Senthil:2003eed}
T.~Senthil, A.~Vishwanath, L.~Balents, S.~Sachdev and M.~P.~A. Fisher,
  \emph{{Deconfined Quantum Critical Points}},
  \href{http://dx.doi.org/10.1126/science.1091806}{\emph{Science} {\bf 303}
  (2004) 1490}, [\href{https://arxiv.org/abs/cond-mat/0311326}{{\tt
  cond-mat/0311326}}].

\bibitem{Cheng:2022sgb}
M.~Cheng and N.~Seiberg, \emph{{Lieb-Schultz-Mattis, Luttinger, and 't Hooft -
  anomaly matching in lattice systems}},
  \href{http://dx.doi.org/10.21468/SciPostPhys.15.2.051}{\emph{SciPost Phys.}
  {\bf 15} (2023) 051}, [\href{https://arxiv.org/abs/2211.12543}{{\tt
  2211.12543}}].

\bibitem{Murthy:1989ps}
G.~Murthy and S.~Sachdev, \emph{{Action of Hedgehog Instantons in the
  Disordered Phase of the (2+1)-dimensional $CP^{N-1}$ Model}},
  \href{http://dx.doi.org/10.1016/0550-3213(90)90670-9}{\emph{Nucl. Phys. B}
  {\bf 344} (1990) 557}.

\bibitem{Takahashi:2024xxd}
J.~Takahashi, H.~Shao, B.~Zhao, W.~Guo and A.~W. Sandvik, \emph{{SO(5)
  multicriticality in two-dimensional quantum magnets}},
  \href{https://arxiv.org/abs/2405.06607}{{\tt 2405.06607}}.

\bibitem{Liu:2018sww}
Y.~Liu, Z.~Wang, T.~Sato, M.~Hohenadler, C.~Wang, W.~Guo et~al.,
  \emph{{Superconductivity from the Condensation of Topological Defects in a
  Quantum Spin-Hall Insulator}},
  \href{http://dx.doi.org/10.1038/s41467-019-10372-0}{\emph{Nature Commun.}
  {\bf 10} (2019) 2658}, [\href{https://arxiv.org/abs/1811.02583}{{\tt
  1811.02583}}].

\bibitem{Chen:2024ddr}
J.-Y. Chen, \emph{{Instanton Density Operator in Lattice QCD from Higher
  Category Theory}},  \href{https://arxiv.org/abs/2406.06673}{{\tt
  2406.06673}}.

\bibitem{Zhang:2024cjb}
P.~Zhang and J.-Y. Chen, \emph{{An Explicit Categorical Construction of
  Instanton Density in Lattice Yang-Mills Theory}},
  \href{https://arxiv.org/abs/2411.07195}{{\tt 2411.07195}}.

\bibitem{Sulejmanpasic:2019ytl}
T.~Sulejmanpasic and C.~Gattringer, \emph{{Abelian gauge theories on the
  lattice: $\theta$-Terms and compact gauge theory with(out) monopoles}},
  \href{http://dx.doi.org/10.1016/j.nuclphysb.2019.114616}{\emph{Nucl. Phys. B}
  {\bf 943} (2019) 114616}, [\href{https://arxiv.org/abs/1901.02637}{{\tt
  1901.02637}}].

\bibitem{Anosova:2019quw}
M.~Anosova, C.~Gattringer, D.~G\"oschl, T.~Sulejmanpasic and P.~T\"orek,
  \emph{{Topological terms in abelian lattice field theories}},
  \href{http://dx.doi.org/10.22323/1.363.0082}{\emph{PoS} {\bf LATTICE2019}
  (2019) 082}, [\href{https://arxiv.org/abs/1912.11685}{{\tt 1912.11685}}].

\bibitem{Gorantla:2021svj}
P.~Gorantla, H.~T. Lam, N.~Seiberg and S.-H. Shao, \emph{{A modified Villain
  formulation of fractons and other exotic theories}},
  \href{http://dx.doi.org/10.1063/5.0060808}{\emph{J. Math. Phys.} {\bf 62}
  (2021) 102301}, [\href{https://arxiv.org/abs/2103.01257}{{\tt 2103.01257}}].

\bibitem{Anosova:2022cjm}
M.~Anosova, C.~Gattringer and T.~Sulejmanpasic, \emph{{Self-dual U(1) lattice
  field theory with a \ensuremath{\theta}-term}},
  \href{http://dx.doi.org/10.1007/JHEP04(2022)120}{\emph{JHEP} {\bf 04} (2022)
  120}, [\href{https://arxiv.org/abs/2201.09468}{{\tt 2201.09468}}].

\bibitem{Anosova:2022yqx}
M.~Anosova, C.~Gattringer, N.~Iqbal and T.~Sulejmanpasic, \emph{{Phase
  structure of self-dual lattice gauge theories in 4d}},
  \href{http://dx.doi.org/10.1007/JHEP06(2022)149}{\emph{JHEP} {\bf 06} (2022)
  149}, [\href{https://arxiv.org/abs/2203.14774}{{\tt 2203.14774}}].

\bibitem{Fazza:2022fss}
L.~Fazza and T.~Sulejmanpasic, \emph{{Lattice quantum Villain Hamiltonians:
  compact scalars, U(1) gauge theories, fracton models and quantum Ising model
  dualities}}, \href{http://dx.doi.org/10.1007/JHEP05(2023)017}{\emph{JHEP}
  {\bf 05} (2023) 017}, [\href{https://arxiv.org/abs/2211.13047}{{\tt
  2211.13047}}].

\bibitem{Jacobson:2023cmr}
T.~Jacobson and T.~Sulejmanpasic, \emph{{Modified Villain formulation of
  Abelian Chern-Simons theory}},
  \href{http://dx.doi.org/10.1103/PhysRevD.107.125017}{\emph{Phys. Rev. D} {\bf
  107} (2023) 125017}, [\href{https://arxiv.org/abs/2303.06160}{{\tt
  2303.06160}}].

\bibitem{Nguyen:2024ikq}
M.~Nguyen, T.~Sulejmanpasic and M.~\"Unsal, \emph{{Phases of theories with
  $\mathbb{Z}_N$ 1-form symmetry and the roles of center vortices and magnetic
  monopoles}},  \href{https://arxiv.org/abs/2401.04800}{{\tt 2401.04800}}.

\bibitem{Jacobson:2024hov}
T.~Jacobson and T.~Sulejmanpasic, \emph{{Canonical quantization of lattice
  Chern-Simons theory}},
  \href{http://dx.doi.org/10.1007/JHEP11(2024)087}{\emph{JHEP} {\bf 11} (2024)
  087}, [\href{https://arxiv.org/abs/2401.09597}{{\tt 2401.09597}}].

\bibitem{Xu:2024hyo}
Z.-A. Xu and J.-Y. Chen, \emph{{Lattice Chern-Simons-Maxwell Theory and its
  Chirality}},  \href{https://arxiv.org/abs/2410.11034}{{\tt 2410.11034}}.

\bibitem{Peng:2024xbl}
C.~Peng, M.~C. Diamantini, L.~Funcke, S.~M.~A. Hassan, K.~Jansen, S.~K\"uhn
  et~al., \emph{{Hamiltonian Lattice Formulation of Compact
  Maxwell-Chern-Simons Theory}},  \href{https://arxiv.org/abs/2407.20225}{{\tt
  2407.20225}}.

\bibitem{Nakayama:2016jhq}
Y.~Nakayama and T.~Ohtsuki, \emph{{Conformal Bootstrap Dashing Hopes of
  Emergent Symmetry}},
  \href{http://dx.doi.org/10.1103/PhysRevLett.117.131601}{\emph{Phys. Rev.
  Lett.} {\bf 117} (2016) 131601},
  [\href{https://arxiv.org/abs/1602.07295}{{\tt 1602.07295}}].

\end{thebibliography}\endgroup

\end{document}